\documentclass[aps,prc,twocolumn,superscriptaddress,floatfix,amsmath,amssymb,nofootinbib,longbibliography]{revtex4-2}
\usepackage[utf8]{inputenc}
\usepackage[T1]{fontenc}

\usepackage{graphicx}
\usepackage{xcolor}
\usepackage{verbatim}
\usepackage{tikz}
\usepackage{dcolumn}

\definecolor{mplblue}{RGB}{0,0,255}
\definecolor{mplred}{RGB}{255,0,0}
\definecolor{mplcyan}{RGB}{0,255,255}
\definecolor{mplgreen}{RGB}{0,128,0}
\definecolor{mplmagenta}{RGB}{255,0,255}
\definecolor{mplorange}{RGB}{255,165,0}

\usepackage[colorlinks=true,citecolor=blue,linkcolor=blue,urlcolor=blue]{hyperref}
\usepackage{orcidlink}

\newcommand{\mev}{\, \text{MeV}}

\newcommand{\isotope}[2]{$^{#2}\text{#1}$}
\newcommand{\chiralorder}[1]{$\text{N}^{#1}\text{LO}$}

\begin{document}

\title{Exploring quark mass dependent three-nucleon forces in medium-mass nuclei}

\author{Urban Vernik\orcidlink{0000-0002-3717-945X}}
\email{urban.vernik@tu-darmstadt.de}
\affiliation{Technische Universit\"at Darmstadt, Department of Physics, D-64289 Darmstadt, Germany} 
\affiliation{ExtreMe Matter Institute EMMI, GSI Helmholtzzentrum f\"ur Schwerionenforschung GmbH, D-64291 Darmstadt, Germany}

\author{Kai Hebeler\orcidlink{0000-0003-0640-1801}}
\email{kai.hebeler@tu-darmstadt.de}
\affiliation{Technische Universit\"at Darmstadt, Department of Physics, D-64289 Darmstadt, Germany} 
\affiliation{ExtreMe Matter Institute EMMI, GSI Helmholtzzentrum f\"ur Schwerionenforschung GmbH, D-64291 Darmstadt, Germany}
\affiliation{Max-Planck-Institut f\"ur Kernphysik, Saupfercheckweg 1, D-69117 Heidelberg, Germany}

\author{Achim Schwenk\orcidlink{0000-0001-8027-4076}}
\email{schwenk@physik.tu-darmstadt.de}
\affiliation{Technische Universit\"at Darmstadt, Department of Physics, D-64289 Darmstadt, Germany} 
\affiliation{ExtreMe Matter Institute EMMI, GSI Helmholtzzentrum f\"ur Schwerionenforschung GmbH, D-64291 Darmstadt, Germany}
\affiliation{Max-Planck-Institut f\"ur Kernphysik, Saupfercheckweg 1, D-69117 Heidelberg, Germany}

\begin{abstract}
Recently, new quark mass dependent three-nucleon (3N) forces have been identified, whose contributions in nuclear matter exceed expectations of Weinberg power-counting arguments. In this work, we investigate the impact of the most dominant new interaction term, characterized by the coupling $F_2$, in {\it ab initio} calculations of medium-mass nuclei. For this, we combine the new $F_2$ interaction with established 3N interactions up to next-to-next-to-leading order (N$^2$LO) and next-to-next-to-next-to-leading order (N$^3$LO) in chiral effective field theory. We explore two fit strategies for the low-energy couplings. The first is based only on few-body observables, while the second also incorporates information from $^{16}$O. Generally, we find that the $F_2$ interaction has a significant impact on energies and radii, however mainly due to changes in the short-range couplings. Overall, we find no significant systematic improvements in the reproduction of medium-mass nuclei when the additional $F_2$ interaction is included.
\end{abstract}

\maketitle

\section{Introduction}

The development of chiral effective field theory (EFT) for nuclear forces~\cite{RevModPhys.81.1773,MACHLEIDT20111,Hammer:2019poc} has been key for the success of {\it ab initio} calculations of nuclei and nuclear matter~\cite{10.3389/fphy.2020.00379,HEBELER20211}. At the same time, powerful many-body advances (see, e.g.,~\cite{BARRETT2013131,Hagen_2014,RevModPhys.87.1067,HERGERT2016165,Barbieri2017,annurev:/content/journals/10.1146/annurev-nucl-101917-021120, PhysRevC.101.014318,tichai2020many}) and systematic uncertainty quantification (see e.g.,~\cite{Furnstahl:2015rha,PhysRevC.100.044001,Huther:2019ont,Hu2022,Maris:2020qne,svensson2025bayesian}) has greatly extended the reach of {\it ab initio} calculations to heavy nuclei~\cite{Stroberg:2019bch,Tichai:2023epe,Hu2022,Miyagi:2023zvv,arthuis2024neutronrichnucleineutronskins}.

In chiral EFT, each contribution to nucleon-nucleon (NN), three-nucleon (3N), and higher-body forces in the low-energy expansion can be assigned an order, which is determined by the underlying power counting. The most widely employed power counting scheme was first formulated by Weinberg~\cite{WEINBERG1990288,WEINBERG19913}.
Recently, Cirigliano {\it et al.}~\cite{kn79-f5m9} studied new quark mass dependent 3N forces with the associated low-energy constants (LECs) $F_2$ and $D_2$. Their analysis indicates that particularly the $F_2$ term provides significant contributions to the energy per particle of nuclear matter, especially to neutron matter. In fact, the magnitude of its contribution appears comparable to and even exceeds those of the leading 3N interactions. This could suggest that $F_2$ might need to be promoted to a lower order than previously assigned in Weinberg power counting. We note that the arguments of \cite{kn79-f5m9} are based on the power counting scheme proposed by Kaplan, Savage, and Wise (KSW)~\cite{Kaplan:1996xu}, and hence it is unclear if the same applies in Weinberg power counting. In addition, the KSW scheme implicitly assumes dimensional regularization of all interactions. When using spectral-function regularization for the $F_2$ term, a linear divergence as a function of the cutoff scale is observed~\cite{private_comm}. Hence, if one decides to promote this 3N interaction to \chiralorder{3}, then also short-range 3N couplings at \chiralorder{4} need to be promoted to the same order to act as counter terms, see also~\cite{Epelbaum:2025wtj}. With these caveats in mind, we will focus in this work on the new interaction terms derived using dimensional regularization and explore their impact on the properties of medium-mass nuclei.

We focus in this work only on the most dominant new interaction term, characterized by the coupling $F_2$, whose value was suggested to be in the range $|F_2| \le 1/5 F_\pi^4$~\cite{kn79-f5m9}. We investigate the impact of the new $F_2$ interaction in medium‑mass nuclei and assess whether its inclusion leads to an improved description of nuclei compared to established state-of-the-art Hamiltonians. To determine the 3N LECs $c_{D}$, $c_{E}$, and $F_{2}$, we employ two different fitting strategies. The first strategy relies exclusively on few‑body observables, whereas the second incorporates the experimental ground‑state energy and charge radius of \isotope{O}{16} in addition to the \isotope{H}{3} binding energy. 

This paper is organized as follows: In Sec.~\ref{sec:calcs}, we present the interaction(s) used in this work along with the calculational details. In Sec.~\ref{sec:strats} two 3N fitting strategies are introduced and their key features discussed. In Sec.~\ref{sec:results} the results for medium-mass nuclei using the new interactions are presented. Finally, we conclude in Sec.~\ref{sec:conclusion}.

\section{Calculational details}
\label{sec:calcs}

In this work we employ the Entem, Machleidt, Noysk (EMN) 450 interaction~\cite{PhysRevC.96.024004} at \chiralorder{2} and \chiralorder{3}. At both chiral orders we consider both evolved and bare interactions. The bare NN interactions are supplemented by contributions from 3N interactions up the corresponding chiral order using a cutoff scale of $\Lambda = 450$\,MeV with the regulator function $f(p,q) = \exp[ - ((p^2 + \frac{3}{4} q^2)/\Lambda_{\text{3N}}^2)^{4}]$. For the evolved interactions, we follow the strategy of~\cite{PhysRevC.83.031301}, and evolve the NN interaction to the resolution scale $\lambda_{\text{SRG}} = 1.8\,\text{fm}^{-1}$ via the similarity renormalization group (SRG). The 3N interactions remain unevolved using a cutoff scale $\Lambda_{3\text{N}} = 2.0 \,\text{fm}^{-1}$. The long-range 3N LECs $c_1$, $c_3$ and $c_4$ are taken from the Roy-Steiner analysis of $\pi N$ scattering \cite{HOFERICHTER20161}, while the intermediate- and short-range LECs $c_D$ and $c_E$ along with $F_2$ are fit according to the strategies discussed in Sec.~\ref{sec:strats}. The summary of interactions used in this work is given in Table~\ref{tab:interactions}.

The interaction matrix elements used for calculations are generated in the spherical harmonic oscillator (HO) basis using the \verb|NuHamil| code~\cite{miyagi2023nuhamil}. Many-body calculations are performed using the in-medium similarity renormalization group (IMSRG) \cite{PhysRevLett.106.222502, HERGERT2016165} at the IMSRG(2) level using the \verb|IMSRG++| code \cite{stroberg2018imsrg++}. The size of the spherical HO basis was set to $e_{\text{max}} = 14$. Unless otherwise specified, a HO frequency $\hbar\omega = 16 \mev$ was employed, and the 3N matrix elements are represented in a basis defined by an additional truncation $E_{3\text{max}} = 24$. 

\section{Fitting strategies}
\label{sec:strats}

\begin{figure}[t!]
    \centering
    \includegraphics[width=\linewidth]{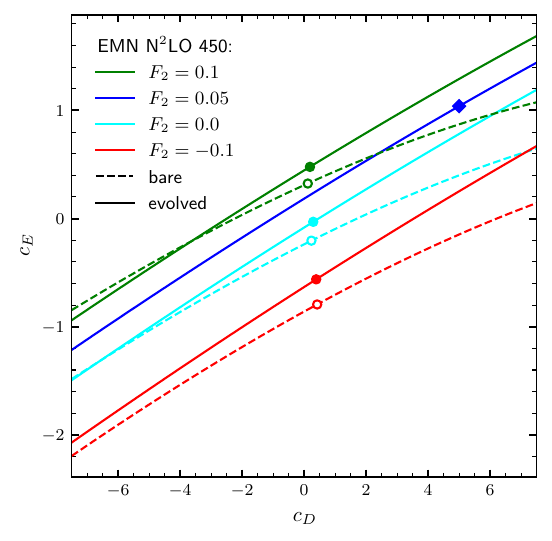}
    \caption{Combinations of 3N LECs $c_D$ and $c_E$ that reproduce the \isotope{H}{3} binding energy for the EMN \chiralorder{2} 450 MeV interaction for different values of $F_2$. Dashed lines correspond to unevolved Hamiltonians and solid lines to low-resolution NN+3N
    interactions at the NN SRG resolution scale $\lambda_{\text{SRG}}=1.8$\,fm$^{-1}$. The symbols indicate the fitted values using $^3$H half-life (circles) and $^{16}$O binding energy plus charge radius (diamond). See main text and Table \ref{tab:interactions} for details.}
    \label{fig:GT_fit_N2LO}
\end{figure}

Given that the long-range 3N couplings $c_i$ are fixed by $\pi N$ scattering,
this leaves two 3N LECs, $c_D$ and $c_E$, up to \chiralorder{3} that need to be constrained using three- or many-body data. With the introduction of $F_2$, in total three LECs need to be fitted. Based on the estimated size of $F_2$ contributions given by Cirigliano {\it et al.}~\cite{kn79-f5m9}, we here consider the range $|F_2| \leq 0.15$. Note that throughout this work we define the value of $F_2$ in units of $F_{\pi}^{-4}$. Furthermore, we explore two different fitting strategies. In both, the values of these three LECs are constrained to reproduce the \isotope{H}{3} binding energy. In Fig.~\ref{fig:GT_fit_N2LO}, we show the resulting dependencies between $c_D$ and $c_E$ at \chiralorder{2} for four different values of $F_2$. The results show that the shape of these lines is essentially independent of $F_2$. In addition, Fig.~\ref{fig:GT_fit_N2LO} demonstrates that changes in $F_2$ and $c_E$ are highly correlated. Specifically, for a given $c_D$ value, positive (negative) values of $F_2$ lead to linear shifts of $c_E$ towards larger, resp.~less negative (smaller, resp.~more negative) values.

\subsection{Fits to experimental \isotope{H}{3} half-life}

The first strategy relies only on few-body observables. Specifically, we constrain the $c_D$ value to reproduce the experimental $^3$H half-life. For the calculation of the Gamow-Teller (GT) matrix elements, we follow the conventions of~\cite{PhysRevLett.103.102502,Klos2017} and employ a non-local regulator to the current matrix elements using a cutoff scale $\Lambda = \Lambda_\text{3N}$. The resulting fit values at \chiralorder{2} for different $F_2$ values are given in the first six rows of Table~\ref{tab:interactions} and are also indicated by circles in Fig.~\ref{fig:GT_fit_N2LO}. This strategy, labeled by ``GT'', favors small values of $c_D$ that exhibit very weak sensitivity on $F_2$ and the SRG resolution scale. In contrast, as stated above, $c_E$ is much more sensitive to~$F_2$.

\renewcommand{\arraystretch}{1.25}
\begin{table}[t!]
    \centering
    \begin{tabular}{c|c|c|c|
    D{.}{.}{2}|   
    D{.}{.}{3}|   
    D{.}{.}{3}|   
    c}
        Label & Order & $\lambda_{\text{SRG}}$ & $\Lambda_{3\text{N}}$ & \multicolumn{1}{c|}{$F_2$}
        & \multicolumn{1}{c|}{$c_D$} & \multicolumn{1}{c|}{$c_E$} & Fit\\
        \hline\hline
        \tikz \draw[fill=white, draw=mplred, line width=1.5pt] (0,0) circle (0.1cm); & \chiralorder{2} & $\infty$ & 450\,MeV & -0.10 & 0.421 & -0.794 & GT\\
        \tikz \draw[fill=white, draw=mplcyan, line width=1.5pt] (0,0) circle (0.1cm); & \chiralorder{2} & $\infty$ & 450\,MeV & 0.00 & 0.235 & -0.204 & GT\\
        \tikz \draw[fill=white, draw=mplgreen, line width=1.5pt] (0,0) circle (0.1cm);& \chiralorder{2} & $\infty$ & 450\,MeV & 0.10 & 0.118 & 0.324 & GT\\
        \hline
        \tikz \draw[fill=mplred, draw=mplred, line width=1.5pt] (0,0) circle (0.1cm); & \chiralorder{2} & 1.8$\,\text{fm}^{-1}$ & $2.0\,\text{fm}^{-1}$ & -0.10 & 0.389 & -0.562 & GT\\
        \tikz \draw[fill=mplcyan, draw=mplcyan, line width=1.5pt] (0,0) circle (0.1cm); & \chiralorder{2} & 1.8$\,\text{fm}^{-1}$ & 2.0$\,\text{fm}^{-1}$ & 0.00 & 0.294 & -0.030 & GT\\
        \tikz \draw[fill=mplgreen, draw=mplgreen, line width=1.5pt] (0,0) circle (0.1cm);& \chiralorder{2} & 1.8$\,\text{fm}^{-1}$ & 2.0$\,\text{fm}^{-1}$ & 0.10 & 0.188 & 0.478 & GT\\
        \hline\hline
        \tikz \draw[fill=white, draw=mplblue, line width=1.5pt]
        (0,0.10cm) -- (0.10cm,0) -- (0,-0.10cm) -- (-0.10cm,0) -- cycle; &  \chiralorder{2} & $\infty$ & 450\,MeV & 0.15 & 7.000 & 1.230 & \isotope{O}{16} \\
        \tikz \draw[fill=white, draw=mplmagenta, line width=1.5pt]
        (0,0.10cm) -- (0.10cm,0) -- (0,-0.10cm) -- (-0.10cm,0) -- cycle; & \chiralorder{3} &  $\infty$ & 450\,MeV & 0.15 & 7.000 & 0.445 & \isotope{O}{16}\\
        \tikz \draw[fill=mplblue, draw=mplblue, line width=0.4pt]
        (0,0.12cm) -- (0.12cm,0) -- (0,-0.12cm) -- (-0.12cm,0) -- cycle; &  \chiralorder{2} & 1.8$\,\text{fm}^{-1}$ & $2.0\,\text{fm}^{-1}$ & 0.05 & 5.000 & 1.039 & \isotope{O}{16} \\
        \tikz \draw[fill=mplmagenta, draw=mplmagenta, line width=0.4pt]
        (0,0.12cm) -- (0.12cm,0) -- (0,-0.12cm) -- (-0.12cm,0) -- cycle; & \chiralorder{3} &  1.8$\,\text{fm}^{-1}$ & $2.0\,\text{fm}^{-1}$ & 0.05 & 5.000 & -0.017 & \isotope{O}{16}\\
        \hline\hline
        \tikz \draw[line width=1.5pt, draw=mplorange] (-0.09cm,-0.09cm) -- (0.09cm,0.09cm) (-0.09cm,0.09cm) -- (0.09cm,-0.09cm); & \chiralorder{2} & 1.8$\,\text{fm}^{-1}$ & $2.0\,\text{fm}^{-1}$ & 0.00 & 5.000 & 1.039 & \isotope{O}{16}\\
        \tikz \draw[line width=1.5pt, draw=mplgreen] (-0.09cm,-0.09cm) -- (0.09cm,0.09cm) (-0.09cm,0.09cm) -- (0.09cm,-0.09cm); & \chiralorder{3} & 1.8$\,\text{fm}^{-1}$ & $2.0\,\text{fm}^{-1}$ & 0.00 & 5.000 & -0.017 & \isotope{O}{16}
    \end{tabular}
    \caption{
    Details on the fitted EMN 450 interactions used in this work, including the corresponding 3N LECs, cutoff and SRG scales, chiral orders, fit strategies and associated plot labels. ``GT'' and ``\isotope{O}{16}'' correspond to the $^3$H half-life and \isotope{O}{16} fitting strategies, respectively. Values of $F_2$ are given in units of $F_{\pi}^{-4}$.}
    \label{tab:interactions}
\end{table}

\begin{figure}[t!]
    \centering
    \includegraphics[width=\linewidth]{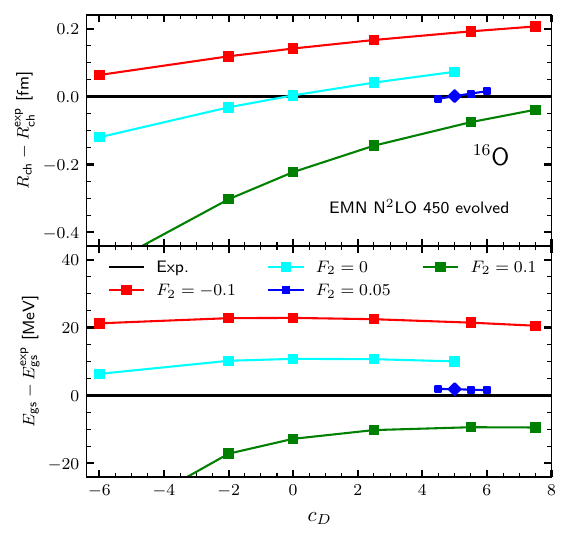}
    \caption{Difference of the calculated $^{16}$O charge radius (top panel) and ground-state energy (bottom panel) to the experimental values as a function of $c_D$ for the EMN \chiralorder{2} 450 evolved interaction at different values of $F_2$. The blue diamond indicates the $c_D$ value that simultaneously reproduces both observables to good approximation for $F_2=0.05$, see also Fig.~\ref{fig:GT_fit_N2LO} and Table \ref{tab:interactions}.}
    \label{fig:cD_scan_general}
\end{figure}

\subsection{Fits to experimental \isotope{O}{16} ground-state energy and charge radius}

The second strategy uses many-body observables as a constraint for the LECs, namely the ground-state energy and charge radius of \isotope{O}{16}. We refer to this strategy as ``\isotope{O}{16}''. Figure~\ref{fig:cD_scan_general} shows the two observables for \isotope{O}{16} based on the evolved interactions as a function of $c_D$. We observe that for increasingly positive $F_2$ values, the ground-state energy and charge radius decrease. The opposite is true in the regime $F_2 < 0$, although the sensitivity of the observables is somewhat smaller. The interaction with $F_2=0$ reproduces the charge radius well at $c_D=0$, while the ground-state energy differs by about 10\,MeV from the experimental value. A value of $F_2 = 0.05$ allows a simultaneous good reproduction of both the ground-state energy as well as the charge radius for $c_D = 5.0$. The corresponding interaction is marked by a blue diamond. However, it is worth noting that the $c_D$ value is only loosely constrained. Within chiral uncertainties, all blue points for this $F_2$ value shown in Fig.~\ref{fig:cD_scan_general} reproduce the experimental value very well.

\begin{figure}[t!]
    \centering
    \includegraphics[width=\linewidth]{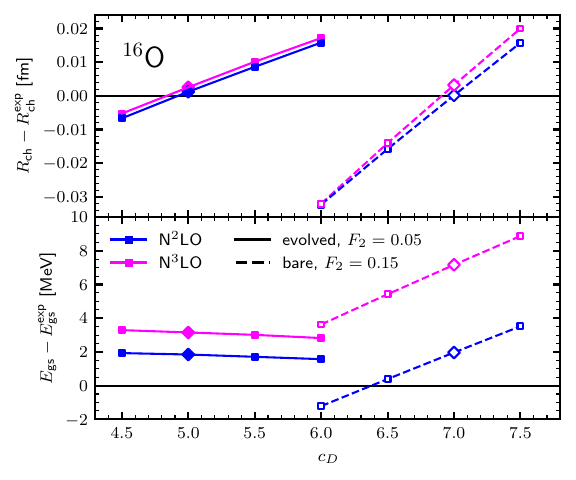}
    \caption{Difference of the calculated $^{16}$O charge radius (top panel) and ground-state energy (bottom panel) to the experimental values as a function of $c_D$ for the EMN 450 evolved (solid lines) and bare interactions (dashed lines) at \chiralorder{2} and \chiralorder{3}. Note the small plot scale for both observables. The diamonds again indicate the fitted interactions listed in Table~\ref{tab:interactions} and used in Sec.~\ref{sec:results}.}
    \label{fig:Strat2_scan}
\end{figure}

Figure~\ref{fig:Strat2_scan} shows the corresponding fits for both chiral orders \chiralorder{2} and \chiralorder{3}, as well as for bare interactions. In all cases, it is possible to obtain very good simultaneous fits of both $^{16}$O observables. While the results at both chiral orders are very similar, the effect of the SRG evolution is significantly larger. We find that the optimal $F_2$ value for bare interactions is about $F_2 = 0.15$ compared to $F_2 = 0.05$ for evolved interactions. The best fit for bare interactions is achieved for $c_D = 7.0$, again much larger than the $c_D$ values obtained using the ``GT'' strategy and in line with the results found in~\cite{arthuis2024neutronrichnucleineutronskins}. However, note again that the $c_D$ value is only loosely constrained, as we find good agreement with experiment for both observables over a range of $c_D$ values.

\subsection{Impact of $F_2$ 3N interaction on nuclei}

\begin{figure}[t!]
    \centering
    \includegraphics[width=\linewidth]{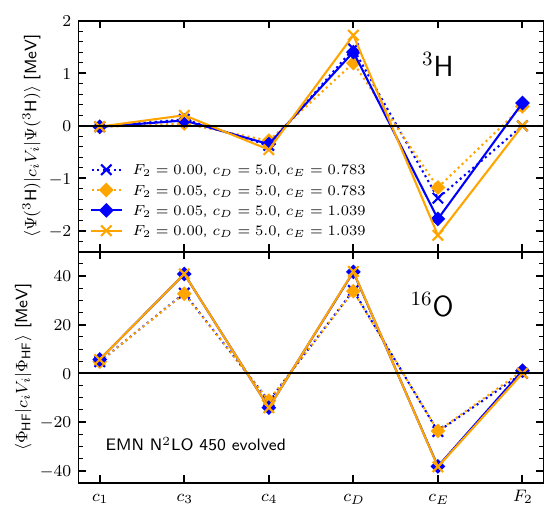}
    \caption{Decomposition of the exact \isotope{H}{3} and Hartree-Fock \isotope{O}{16} ground-state energies into individual 3N and $F_2$ contributions at N$^2$LO. The blue crosses show the results of a reference interaction with $F_2 = 0$, $c_D = 5.0$, and the $c_E$ value fitted to the $^3$H binding energy. The orange diamonds correspond to an interaction with $F_2 = 0.05$ added, but without refitting $c_E$. The blue diamonds represent the interaction where $c_D$, $c_E$, and $F_2$ are refitted consistently using the ``\isotope{O}{16}'' strategy, and orange crosses keep the refitted $c_D$ and $c_E$, but with $F_2 = 0$. The \isotope{H}{3} expectation values were obtained from solutions of the Faddeev equations. For the Hartree-Fock calculations of \isotope{O}{16}, $E_{3\text{max}} = 16$ was used.}
    \label{fig:HFDecomposition}
\end{figure}

In order to gain deeper insight into the impact of the $F_2$ interaction on properties of nuclei, it is instructive to study the contributions of the individual 3N interactions. In Fig.~\ref{fig:HFDecomposition} we show the expectation values of the N$^2$LO contributions plus the $F_2$ term for the full $^3$H wave functions (upper panel) and the $^{16}$O Hartree-Fock wave function (lower panel), based on the evolved interaction. For \isotope{O}{16}, Hartree-Fock results have the advantage that the energy contributions are linear in the different LECs and hence there are no interference effects between the different contributions like, e.g., in the IMSRG(2) solution. This allows a cleaner analysis of the underlying mechanisms, even though the Hartree-Fock energy can differ quite significantly from the full IMSRG(2) result. As a reference, we show the contributions for an interaction without $F_2$ using the same $c_D$ like the evolved interaction at N$^2$LO fitted to $^{16}$O observables, whereas the $c_E$ value is determined from a fit to the $^3$H binding energy (blue crosses). By definition, the expectation value of the $F_2$ term is exactly zero in this case. Adding the $F_2$ interaction and leaving all other LECs invariant (orange diamonds), has a notable impact on $c_D$ and $c_E$ expectation values in \isotope{H}{3}, but only negligible effects in \isotope{O}{16}. This implies, first, that the $F_2$ term for $F_2 = 0.05$ induces much larger changes in the wave function for lighter \isotope{H}{3} than for heavier \isotope{O}{16}, where the Hartree-Fock wave function is only marginally affected. Second, it shows that the expectation value of the $F_2$ interaction itself for light systems can be significant, while for heavier isotopes it is very small. In total, for \isotope{O}{16}, the Hartree-Fock energy only increases by about $0.9$\,MeV. These results are compatible with the trends of symmetric nuclear matter Hartree-Fock results shown in~\cite{kn79-f5m9} at nuclear densities. On the other hand, when adding the $F_2$ term using $F_2 = 0.05$, but now refitting the $c_E$ LEC to the $^3$H binding energy (blue diamonds), we find significantly larger changes in the individual contributions in heavier nuclei like \isotope{O}{16}, and a total decrease of the Hartree-Fock energy of about $8.6$\,MeV. Finally, when setting $F_2 = 0$ while retaining the refitted values of $c_D$ and $c_E$ (orange crosses), we again observe significant changes in the \isotope{H}{3} energy and almost no change in \isotope{O}{16} compared to the case including $F_2$. This interaction overbinds \isotope{H}{3} by $\sim0.5$\, MeV at both chiral orders. Altogether, this demonstrates that the effect of the $F_2$ interaction is more prominent in light systems like $^3$H, as seen explicitly in Fig.~\ref{fig:GT_fit_N2LO}. This implies that the main effect of $F_2$ is to impact the few-body fits, which modifies the value of the short-range LECs like $c_E$, which in turn can have a significant impact on observables of heavier nuclei. The expectation value of $F_2$ itself, however, always remains comparatively small in our studies.

\section{Results}
\label{sec:results}

\begin{figure}[t!]
    \centering
    \includegraphics[width=\linewidth]{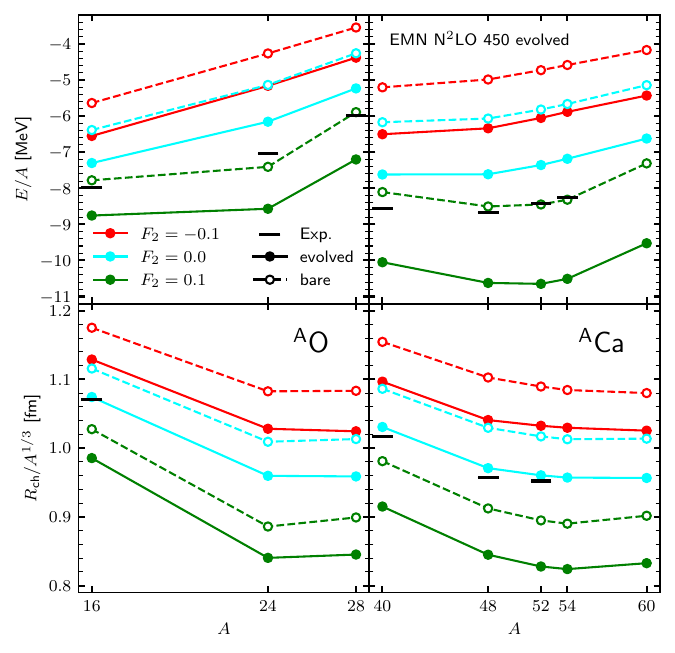}
    \caption{Ground-state energy per particle (top panels) and charge radius (bottom panels) of closed-shell oxygen (left) and calcium (right) isotopes based on bare and evolved EMN \chiralorder{2} 450 interactions for different $F_2$ values, following the ``GT'' fitting strategy (see Table~\ref{tab:interactions}). Experimental data for ground-state energies are taken from~\cite{Wang_2021}, \isotope{O}{16} radius from~\cite{ANGELI201369}, and calcium radii from~\cite{GarciaRuiz2016}.}
    \label{fig:GT_F2scan}
\end{figure}

Next, we study the interactions discussed in the previous section in IMSRG(2) calculations of medium-mass nuclei. We particularly focus on the question of whether the addition of the new $F_2$ interaction leads to a significant improvement.

\begin{figure*}[t!]
    \centering
    \includegraphics[width=0.8\linewidth]{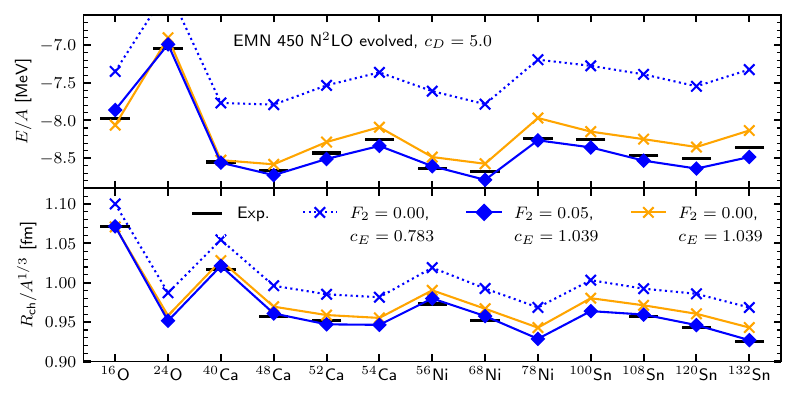}
    \caption{Ground-state energies and charge radii for medium-mass doubly closed-shell nuclei, compared to experiment for the interactions with $F_2 = 0$ (crosses) and $F_2 = 0.05$ (diamonds) at N$^2$LO. For the latter, the ``\isotope{O}{16}'' fitting strategy was used (see Table \ref{tab:interactions}). Orange crosses show results for the refitted interaction, with $F_2=0$.} Experimental data for ground-state energies are taken from~\cite{Wang_2021}, calcium radii from~\cite{GarciaRuiz2016}, \isotope{Ni}{56} from~\cite{PhysRevLett.129.132501}, \isotope{Ni}{68} from~\cite{PhysRevLett.128.022502}, and all others from~\cite{ANGELI201369}. For Ni and Sn isotopes, $\hbar\omega = 12 \mev$ was used.
    \label{fig:ClosedShells}
\end{figure*}

\begin{figure*}[t!]
    \centering
    \includegraphics[width=0.8\linewidth]{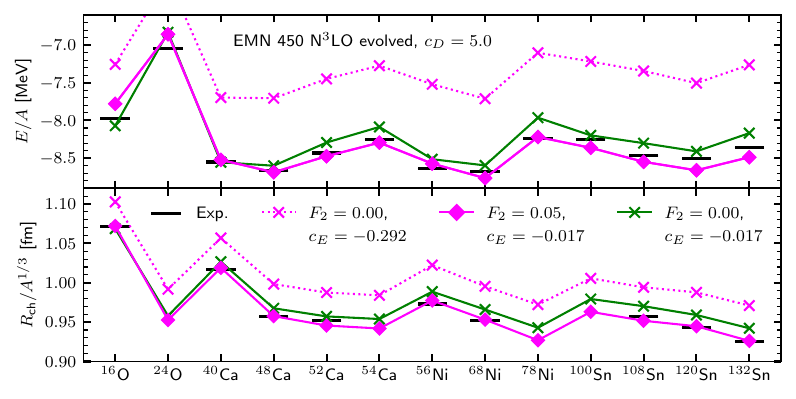}
    \caption{Same as Fig. \ref{fig:ClosedShells} but for \chiralorder{3}.}
    \label{fig:ClosedShellsN3LO}
\end{figure*}

Figure~\ref{fig:GT_F2scan} shows the ground-state energies per particle and charge radii for closed-shell oxygen and calcium isotopes using the ``GT'' strategy, i.e., when using only few-body observables for the interaction fits. For all shown interactions, adding $F_2$ clearly has an important impact on energies and radii. The effect is mostly uniform along a given isotopic chain and scales with mass number, hence the contributions are larger for the calcium chain than for the oxygen isotopes. As already noted above, $F_2 > 0$ generally implies more binding and smaller charge radii. The results also show that none of the interactions can simultaneously reproduce both observables. As shown in Fig.~\ref{fig:cD_scan_general}, variations in $c_D$ as a function of $F_2$ are required to vary the charge radius independently of the energy. Because for the ``GT'' strategy, the value of $c_D$ is essentially independent of $F_2$, the energy and radius will always change in a correlated way, implying that no simultaneous reproduction of both observables is possible for any value of $F_2$.

In contrast, with the fitting strategy ``\isotope{O}{16}'' it is possible to simultaneously reproduce energies and charge radii as shown, e.g., in Fig.~\ref{fig:Strat2_scan}. In Figs.~\ref{fig:ClosedShells} and~\ref{fig:ClosedShellsN3LO} we show the ground-state energies per particle and the scaled charge radii for a selection of doubly closed-shell nuclei across the nuclear chart based on the evolved interactions at orders \chiralorder{2} and \chiralorder{3}. Generally, we find a good reproduction across the full mass range for the interactions including $F_2$ at both chiral orders. The energies are reproduced within $2\%$ ($3\%$) at \chiralorder{2} (\chiralorder{3}). Similarly, the charge radius is also very well reproduced with deviations from experiment below $1\%$. We find that the induced changes by the $F_2$ interaction are larger for heavier isotopes, where they generally provide an improvement in the reproduction of the experimental data compared to the interactions without $F_2$. The contribution of the new interaction, however, always remains smaller than that of $c_E$. As illustrated in Figs.~\ref{fig:ClosedShells} and~\ref{fig:ClosedShellsN3LO}, the ground-state energy and charge radius of lighter nuclei, up to \isotope{Ca}{40}, can be reproduced with comparable accuracy by setting $F_2 = 0$ and keeping the refitted values of $c_D$ and $c_E$. This yields an interaction of similar quality without introducing an additional LEC, at the expense of overbinding the \isotope{H}{3} ground-state energy and small deviations in heavier nuclei (marginal underbinding and slightly overestimated radii). These conclusions are valid at both chiral orders. Moreover, the bare interactions at both chiral orders also work well for light nuclei, but are unable to reproduce the observables past \isotope{Ca}{48}. Beyond this point, the Hartree-Fock ground states become unbound and hence represent a poor reference state for IMSRG calculations.

\begin{figure}
    \centering
    \includegraphics[width=\linewidth]{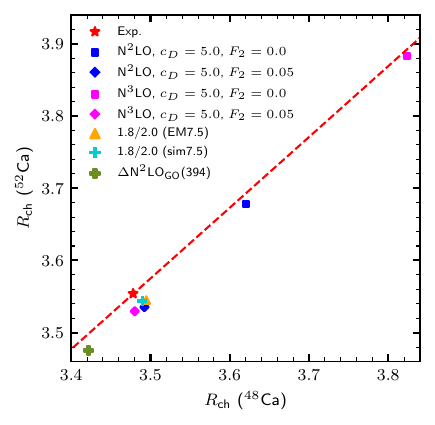}
    \caption{$R_{\text{ch}}(^{52}\text{Ca})$ versus $R_{\text{ch}}(^{48}\text{Ca})$ for the evolved EMN 450 interactions for different $F_2$ values, compared to other established state-of-the-art Hamiltonians~\cite{arthuis2024neutronrichnucleineutronskins, PhysRevC.102.054301} and the experimental value~\cite{GarciaRuiz2016}. The red dashed line corresponds to the experimental value $\delta\langle R_{\text{ch}}^2\rangle^{48,52}_{\text{exp}} = 0.530(5)$\,fm$^2$.}
    \label{fig:4852Rch}
\end{figure}

We also studied the effect of $F_2$ on the prediction of the differential charge radius $\delta\langle R_{\text{ch}}^2\rangle^{48,52} = R_{\text{ch}}^2(^{52}$Ca$) - R_{\text{ch}}^2(^{48}$Ca$)$. Figure~\ref{fig:4852Rch} shows predictions for the charge radius of \isotope{Ca}{52} as a function of the \isotope{Ca}{48} charge radius based on established as well as the new interactions developed in this work. For the fitted evolved interactions we find $\delta\langle R_{\text{ch}}^2\rangle^{48,52}_{\text{N$^2$LO}} = 0.301$\,fm$^2$ and $\delta\langle R_{\text{ch}}^2\rangle^{48,52}_{\text{N$^3$LO}} = 0.346$\,fm$^2$. These values are comparable to the predictions of other Hamiltonians~\cite{arthuis2024neutronrichnucleineutronskins, PhysRevC.102.054301}, but deviate from the experiment even more than for the interaction for which $F_2=0$ was chosen. This shows that the inclusion of the $F_2$ interaction cannot resolve the discrepancy to the experimental value $\delta\langle R_{\text{ch}}^2\rangle^{48,52}_{\text{exp}} = 0.530(5)$\,fm$^2$~\cite{GarciaRuiz2016}.

\section{Conclusions}
\label{sec:conclusion}

In this work, we investigated the impact of new quark mass dependent 3N forces~\cite{kn79-f5m9} on the properties of medium-mass nuclei. To this end, we combined the new $F_2$ interaction with established 3N contributions and the EMN 450 NN interaction, both at N$^2$LO and N$^3$LO. Generally, we find that the $F_2$ term provides significant contributions, especially in very light nuclei like $^3$H. As a consequence, it mainly leads to a change of the short-range 3N couplings when fitting them to few-body observables, which in turn can induce sizable effects on observables of heavier nuclei. While the contribution of $F_2$ increases as function of mass, it always remains small compared to $c_D$ and $c_E$.
This observation holds at both \chiralorder{2} and \chiralorder{3}, and for bare and evolved NN interactions. These findings are consistent with the arguments given in \cite{Epelbaum:2025wtj}, which indicate that adding the $F_2$ term mainly leads to a reparametrization of the interaction, rather than the introduction of new physics. Moreover, we only find a slight systematical improvement in the agreement with experimental energies and radii for heavier doubly closed-shell nuclei compared to the interactions with refitted $c_D$ and $c_E$, but without $F_2$. Furthermore, we find no improved description of the large charge radius increase from $^{48}$Ca to $^{52}$Ca, although $F_2$ offers another fit parameter. In summary, these findings do not provide evidence that would justify promoting $F_2$ to lower order in the chiral expansion in Weinberg power counting.

\begin{acknowledgments}
We thank Pierre Arthuis, Vincenzo Cirigliano, Maria Dawid, Wouter Dekens, Evgeny Epelbaum, Takayuki Miyagi, and Sanjay Reddy for useful discussions. This work was supported in part by the LOEWE Top Professorship LOEWE/4a/519/05.00.002(0014)98 by the State of Hesse. The authors gratefully acknowledge the Gauss Centre for Supercomputing e.V. \cite{gauss} for funding this project by providing computing time through the John von Neumann Institute for Computing (NIC) on the GCS Supercomputer JUWELS at Jülich Supercomputing Centre (JSC).
\end{acknowledgments}

\section*{Data Availability}
The data that support the findings of this article are openly available \cite{vernik_2026_20119597}.

\bibliography{bibliography}

\end{document}